\documentclass[iop]{emulateapj}

\usepackage{natbib}                
\shorttitle{A 1574-DAY PERIODICITY OF TRANSIT(S) ORBITING KIC 8462852}
\shortauthors{Sacco et al.}

\begin{document}


\title{A 1574-DAY PERIODICITY OF TRANSITS ORBITING KIC 8462852}


\author{G. Sacco\altaffilmark{1}, L. Ngo\altaffilmark{2}}
\affil{Citizen Scientists}

\and

\author{J. Modolo\altaffilmark{3}}
\affil{Laboratoire Traitement du Signal et de l'Image, 35042 Rennes, France}

\altaffiltext{1}{Citizen Scientist, gdsacco@hotmail.com.}
\altaffiltext{2}{Citizen Scientist.}
\altaffiltext{3}{INSERM, Rennes 1 University, LTSI, Rennes, F-35000, France.}


\begin{abstract}
Observations of the main sequence F3 V star KIC 8462852 (also known as Boyajian's star) revealed extreme aperiodic dips in flux up to $\sim 20$\% during the four years of the \textit{Kepler} mission. Smaller dips ($<$ 3\%) were also observed with ground-based telescopes between May 2017 and May 2018. We investigated possible correlation between recent dips and the major dips in the last 100 days of the \textit{Kepler} mission. We compared Kepler light curve data, 2017 data from two observatories (TFN, OGG) which are part of the Las Cumbres Observatory (LCO) network, as well as archival data from the Harvard College Observatory (HCO), Sonneberg Observatory and Sternberg Observatory, and determined that observations appear consistent with a 1,574-day (4.31 year) periodicity of a transit (or group of transits) orbiting Boyajian’s star within the habitable zone. Comparison with future observations is required to validate this hypothesis. Furthermore, it is unknown if transits that have produced other major dips as observed during the \textit{Kepler} mission (e.g. D792) share the same orbital period. Nevertheless, the proposed periodicity is a step forward in guiding future observation efforts.
\end{abstract}






\section{Introduction}


To identify exoplanetary transits, the \textit{Kepler} mission measured the brightness of objects within a portion of the sky between Cygnus and Lyrae over a period of approximately four years (2009 to 2013) with a 30-minute cadence. During this observation period, the mission targeted more than 150,000 stars, finding over 2,300 confirmed exoplanetary transits. Citizen scientists in the Planet Hunters program (https://www.planethunters.org/) helped identify KIC 8462852 via its highly unusual and enigmatic light curve. Yet, additional follow-up ground based observations reveal an ordinary main sequence F star with no apparent IR excess. The stars light curve exhibits aperiodic irregularly shaped dips ranging from ~0.2\% to ~22.0\%. It is intriguing to note that a  quasi-periodicity of 24.2 days (between a subset of dips) was identified by Boyajian et al. in 2016, and a 1574-day periodicity is equivalent to 65.0 X 24.2. This Kepler Las Cumbres Observatory (LCO) comparison adds additional support to this Boyajian et al. (2016) finding. In addition, Boyajian et al. detected a 0.88 day periodicity in the Kepler photometric timeseries. They noted that the ~0.88-day signal is likely related to the rotation period of the star ($84\pm4$ km/s), but a paper published by \citet{makarov} suggests this may be due to contamination by another source in the Kepler field. It is debatable as to whether this signal originates from a distant companion star. 

In the present paper, we examined 2017 ground-based observations and data provided by LCO as they compare to the final set of dips observed in 2013 by the Kepler Space Telescope. In addition, we also discuss the possible historical dip detections of October 24, 1978; April 30, 1944; and August 21, 1935. As we detail below, these historical findings align to a 1574.4-day periodicity. 


\section{Observations and Analysis}

    \subsection{Datasets}

\begin{figure}[h]
  \begin{center}
          \includegraphics[width=80mm, height=60mm] 
          {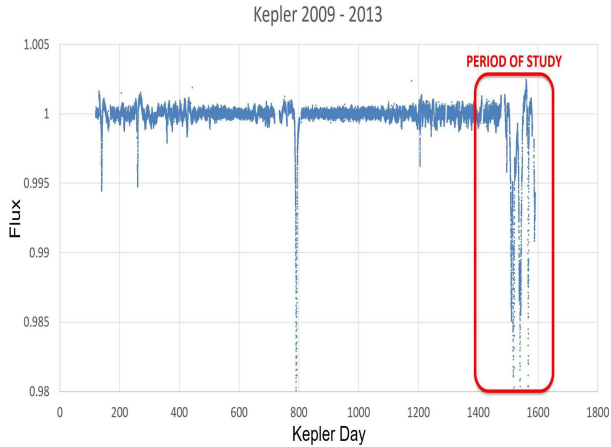}
    \caption{A visual representation of the full \textit{Kepler} light curve for KIC 8462852 (1-May-2009 – 11-May-2013). The period of study includes a range from D1400 - D1590. Lower limit flux range is limited to 0.98 to allow for clearer illustration of all dip events. Several dips drop significantly deeper, for example, D792, D1519, and D1568 drop by 18\%, 22\%, and 8\% respectively.}\label{new_fig_kepler}
  \end{center}
\end{figure}

Two primary sets were adopted for analysis: The four year long cadence {\it Kepler} photometric time-series and observations from the LCO.   First, we used normalized Kepler Space Telescope data containing all 1,471 days that the mission observed KIC 8462852, Data Release 14 (see Figure \ref{new_fig_kepler}). This photometry is based on subrastered imaging, which are made publicly available as soon as calibration is complete. They can be downloaded from a dedicated data retrieval page at Mikulski Archive for Space Telescopes (MAST): http://archive.stsci.edu/kepler/ffi/search.php.  It is important to also note that the Kepler spacecraft transmitted data once per month. Every 3 months the spacecraft was rotating to orient its solar cells towards the Sun. As a result, there are monthly gaps in the observations and a larger gap every three-months when the spacecraft was re-positioned.

\begin{figure}[ht]
  \begin{center}
          \includegraphics[width=80mm, height=60mm] 
          {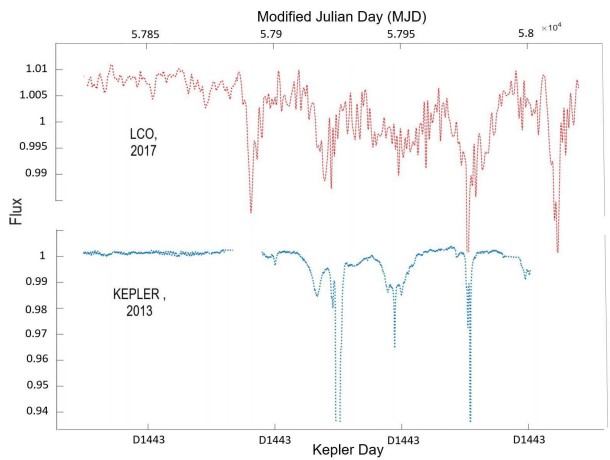}
    \caption{\textit{Kepler}, bottom, light curve for KIC 8462852 (November 2 2012 to May 11 2013) compared to LCO, top, light curve (February 22, 2017 - September 19, 2017) using a 1574-day periodicity. Note that LCO first started observations in February 2017 and recorded no dips prior to Elsie, which is visually consistent to Kepler during the same period. Also note that breaks in Kepler line curve represent missing data due to malfunction, or changing orientation of the space telescope.  LCO data displayed with an overall moving average applied.}\label{zoom_kepler}
  \end{center}
\end{figure}

Second, we used r-band daily averages taken by the LCO 0.4m telescope network as presented in Boyajian et al. (2018). The LCO ground-based observations alerted astronomers starting in May 2017 when a nascent dip was observed, later nick-named “Elsie.” The {\it Elsie} dip was followed by additional dips observed in subsequent LCO observations. For simplicity, we will refer to \textit{Kepler} dips with a "D" followed by the mission day when peak depth was recorded and we will refer to the 2017 dips by their given name as nominated through Kickstarter contributors (see Table 1). An early / mid-July 2017 dip was never named due to its shallow depth. We refer to that dip by the calendar date of peak depth (July 14, 2017) in the remainder of the paper. A comparison of Kepler and LCO data is presented in Figure \ref{zoom_kepler}.

\begin{table}
\begin{center}
\caption{Kepler / LCO Comparison - A comparison of Kepler (2013) and LCO (2017) peak dip dates.  Note the period (days) between each peak. \label{tbl1}}
\begin{tabular}{crrc}
\tableline
\tableline
Dip & Observatory & Peak (MJD) & Period (Days)  \\
\tableline
D1487 & Kepler & 56319 & - \\
Elsie & OGG & 57893 & 1574 \\
D1519 & Kepler & 56351 & - \\
Celeste & TFN & 57925 & 1574.6 \\
D1541 & Kepler & 56373 & - \\
Mid-July & OGG & 57948 & 1575 \\
D1568 & Kepler & 56400 & - \\
Skara Brae & TFN & 57974 & 1574.5 \\
\tableline
\end{tabular}


\end{center}
\end{table}

    \subsection{Quantifying similarity between 2013 and 2017 dips}

In order to quantify the similarity between the dip sequences, which occurred in 2013 (observed by \textit{Kepler}) and in 2017 (observed by the LCO network), we computed different cross-correlograms aiming to identify the periodicity corresponding to an optimal agreement between time-lagged versions of these two signals.

A correlation coefficient measures the extent to which two variables tend to change together. The coefficient describes both the strength and the direction of the relationship. Minitab (http://www.minitab.com/en-US/default.aspx)  offers two different correlation analyses.  Correlation coefficients only measure linear (Pearson) or monotonic (Spearman) relationships. We used both cross-correlograms: 
\begin{itemize}
    \item Linear correlation: The Pearson correlation evaluates the linear relationship between two variables. A relationship is linear when there is a change in one variable that is associated with a proportional change in the other.
    \item Monotonic correlation: The Spearman correlation evaluates the monotonic relationship between two variables. In a monotonic relationship, the variables tend to change together, but not always at a constant rate. This correlation coefficient uses ranked values for each variable.
\end{itemize}

We note that these cross-correlograms were applied to the raw data, without any detrending or normalization.


\section{Results}

    \subsection{Hypothesis}

We produced cross-correlograms of data from the LCO network and \textit{Kepler}. Since the amount of data was not sufficiently large, it was not our intent to use correlation tests to establish statistical significance. We actually used such tests to support our pre-existing goodness of fit hypothesis of 1574 days periodicity, that we found by matching the Kepler and LCO light curves. After performing the correlation, we found three plausible dip matchings, but only one (1572) worked in terms of lining up the Kepler Q4 light curve vs the LCO 2017 light curve. Therefore, these tests supported the original hypothesis. However, statistical significance has not been reached yet, which will need further observational data to reach this benchmark.

In the comparison of Kepler to LCO data, it is worth pointing out the differences in observation frequency between the two.  Kepler data has a higher sampling rate (one point every 29.4 minutes). While LCO used two observatories, the rates are significantly lower due to required night coverage and weather conditions. Since \textit{Kepler} has data gaps that might bias results in favor/against non-dips/dips if interpolated, we skipped any comparisons falling within a \textit{Kepler} gap of half a day or more. The results produced by both methods show three potential hypothesis correlations suggesting a periodicity of: $\approx$ 1540 days, $\approx$ 1572 days and $\approx$ 1600 days.

\begin{figure}[h]
  \begin{center}
          \includegraphics[width=80mm] 
          {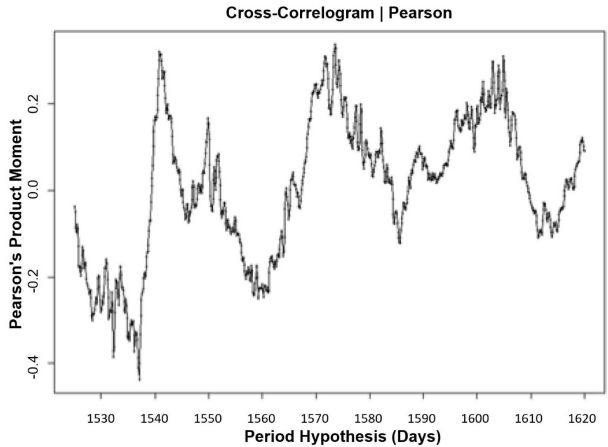}
    \caption{Cross-correlogram between \textit{Kepler} and LCO data based on Pearson's Product Moment. Maximum values suggest a  correlation between both datasets.}\label{pearson}
  \end{center}
\end{figure}

A cross-correlogram based on Pearson's Product Moment is presented in Figure \ref{pearson}. Our three matching hypotheses are depicted in the cross-correlogram, corresponding to periods of ~1540 days, ~1572 and ~1600 days, respectively. Both peaks have similar correlation values; however, the peak corresponding to hypothesis 1 is brief. The peak of hypothesis 2 is broader, suggesting there's greater flexibility in terms of finding a good match and that this periodicity is more robust. A third peak corresponding to $\approx$ 1600 days, while broad, is clearly shallower than the first two hypotheses.

\begin{figure}[h]
  \begin{center}
          \includegraphics[width=80mm] 
          {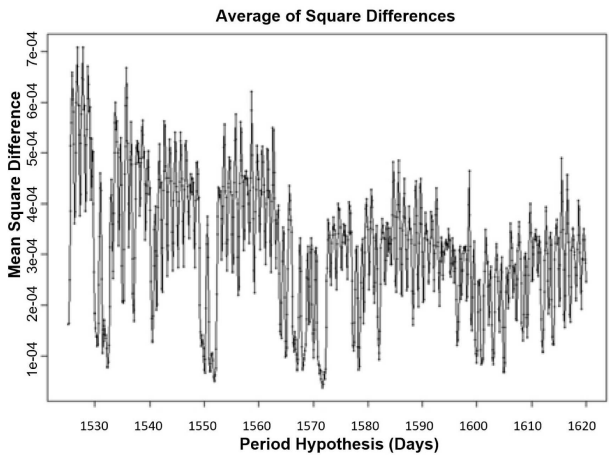}
    \caption{Mean squared error between the \textit{Kepler} and LCO data for different values of hypothetical periodicities. Minimum values suggest a potential correlation between both datasets.}\label{error}
  \end{center}
\end{figure}

Since Pearson's Product Moment is a normalized covariance metric, favoring matching between signals of similar phase and frequency but irrespective of amplitude, we also examined the average square differences between the time series for different period hypotheses (Figure \ref{error}). In this case, we were looking for minima. While the average of square differences is clearly a less stable metric, it also supports our view that hypothesis 2 appears more plausible than hypothesis 1.

\begin{figure}[h]
  \begin{center}
          \includegraphics[width=80mm] 
          {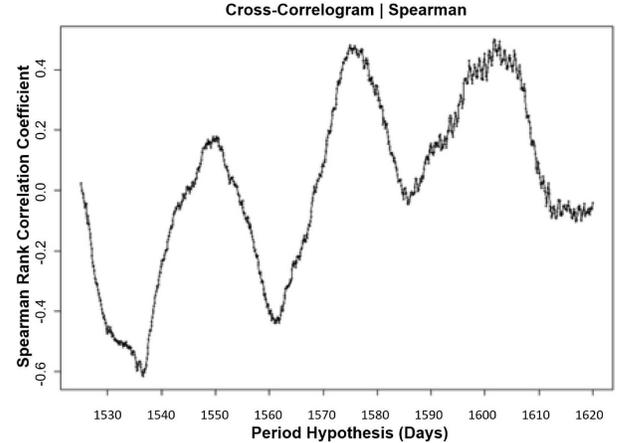}
    \caption{Cross-correlogram computed using Spearman's rank-based correlation. Maximum values suggest a possible correlation between both datasets.}\label{spearman}
  \end{center}
\end{figure}

Finally, and most importantly, we found a rank-based correlation in Spearman's decisively favored hypothesis 2, as shown in Figure \ref{spearman}. A rank-based correlation only considers how well the order of observations matches across both time series, and does not consider flux values beyond their use in sorting observations.

In summary, in two of the three correlation analyses conducted,  hypothesis 2 (~1574 days) had a slightly higher plausibility. Also, the Spearman's rank-based correlation more clearly favors hypothesis 2. In the end, all three methods point to hypothesis 2. 

    \subsection{Hypothesis 2: visual comparison of \textit{Kepler} D1487 to D1590 to LCO May-September 2017}

The final days of \textit{Kepler} (Figure \ref{new_fig_kepler}) are an interesting and active period presenting an intriguing result for this analysis.  Since there is a series of dips within its approximately last 100 days, it provided an ideal visual test to all three hypotheses.  We created three overlay line graphs (1540, 1572, 1600) and found only one with clear visual alignment (1572).  Both hypothesis 1 and 3 failed to align visually, and given this, we discontinued consideration of these two results.

\begin{figure}[h]
  \begin{center}
          \includegraphics[width=85mm, height=65mm] 
          {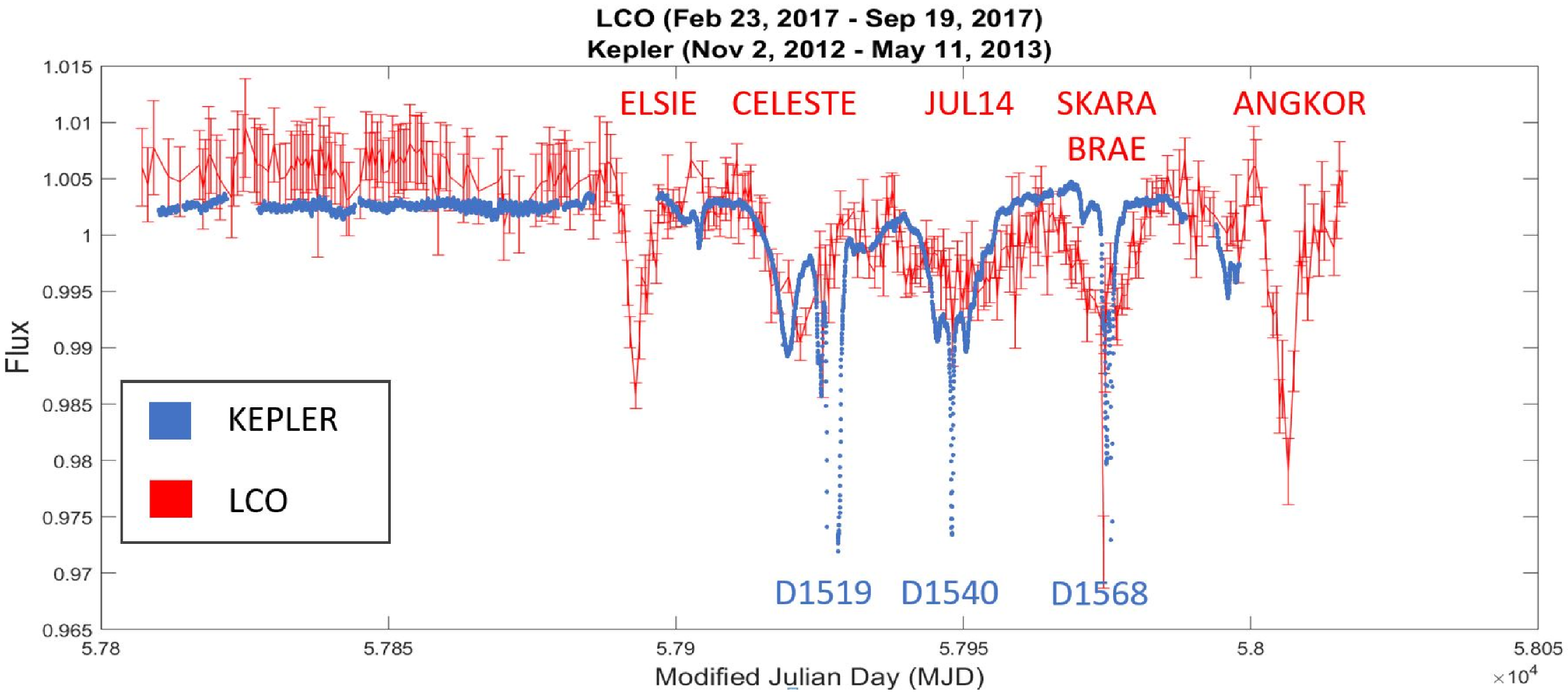}
    \caption{This overlay compares the final Kepler dips in 2013 to the more recent ground-based observed dips of 2017. Using the clear favored hypothesis 2 (1574-day periodicity), we overlaid the Kepler light curve from day 1401 to 1609 with the LCO light curve from 57807 to 58015 (Modified Julian Date). Missing periods of blue (Kepler) light curve are due to lapses of observation from the space telescope.}\label{lco_overlay_kepler}
  \end{center}
\end{figure}

Using the favored hypothesis 2 ($\approx$ 1572-day periodicity), our analysis then focused to more precisely refine this by visually inspecting each light curve comparison overlay.  We performed this analysis and highlighted each result in the following sections.  This review included the same period of \textit{Kepler} days 1401 to 1609 with the LCO light curve from 57807 to 58015 (Modified Julian Date). We find that overall an $\approx$ 1572-day period compares well; however, a slightly refined periodicity of $\approx$ 1574 provides a more precise visual fit.  We examine each dip correlation (\textit{Kepler} vs LCO, see Figure \ref{lco_overlay_kepler}) in the subsequent sections using a 1574-day periodicity. 

The LCO ``Elsie'' observation during May of 2017 is an interesting fit because when you subtract 1574 days, you arrive in the Kepler data during a period in which no observations were being made.  It turns out that over the four years of the \textit{Kepler} mission, observations were interrupted for a variety of reasons. On a regular basis the spacecraft rotated and recalibrated causing a short down-period of observations. In other cases, mechanical failures caused more extended lapses as for example between the \textit{Kepler} period 1477 and 1489 when no observations were made. Based on a 1574-day periodicity, we hypothesize that a dip corresponding to Elsie started on \textit{Kepler} day 1484 and ended on day 1489, but was not observed by \textit{Kepler} (see Figure 6).

We compare the LCO ``Celeste'' dip to the Kepler D1519 dip. We note that there are only about 23 LCO observations to characterize Celeste whereas there are over 900 observations which depict D1519. Yet, even with the limited number of observations, the depression in the light curve, and timing, during this epoch is clear (see Figure 6). 

Next we compared D1540 to the LCO depression that peaked on about July 14, 2017. Both D1540 and the July 14, 2017 event was a complex set of dips and sub-dips (see again Figure 6). These dips were also the shallowest when compared to the other dips highlighted in the present paper.  Ground-based observations in 2017 only detected a maximum depression of ~1.12\%.  However, the timing of dips across what appears to be a complex and lengthy period is correlated. The maximum dip intensity during this period was recorded by LCO on July 14, 2017 and 1574 days earlier, recorded the maximum intensity of that D1540 dip.

On August 9, 2017, the “Skara Brae” dip peaked at almost 3\%. 1574 days prior, Kepler D1568 peaked as well (as can be seen in Figures 6).  Again, there is good agreement in the timing of each event's maximum dip amplitude.  

While a clear matching of duration and peak dip timing between Kepler Q4 and LCO 2017 can be seen, the dip intensity is different. In their paper, \citet{boyajian2018} point out that dip intensity may be expected to change on subsequent orbits if what we are seeing are small-dust particle concentrations. This is because such optically thin dust (with a size scale $< 1$ micrometer) would be quickly blown out of the system. Thus, for each subsequent orbit, the amount of new dust being blown off would likely be different causing a changing depth of stellar dimming.


    \subsection{1574-day period and a look back at \textit{Kepler}}

When we merge \textit{Kepler} with Elsie and Angkor via a 1574-day period, some symmetry is apparent in both overlaid light curves (Figure \ref{lco_overlay_kepler}). Dip D1540, sometimes described as a “triplet,” might be visualized as the centroid of a group of dips. Similarly, 2017 LCO dips might be considered to be an approximately symmetric group with a centroid around July 14, 2017.


    \subsection{Other observations and pre-Elsie comparison to \textit{Kepler}}

There are no reported and/or confirmed dips detected from the end of the \textit{Kepler} mission and prior to Elsie. Data sources for this period include, but are not limited to: AAVSO, LCO, SWIFT, Spitzer, ASAS (or ASAS-SN) and Bruce Gary's\footnote{http://brucegary.net/ts3/} observations. These sources had various start times of regular observations and differing degrees of accuracy. For example, it is unlikely that AAVSO could detect dip intensities lower than 1\%. All these other sources are consistent in that no dips were detected prior to Elsie, which is another factor supporting a 1574-day periodicity. Consequently, it is reasonable to conclude the epoch between August 25, 2016 and May 7, 2017 as consistent with \textit{Kepler} as having nominal flux. 

\section{Testable predictions}

We raise the possibility that a 1574-day periodicity presents opportunities for confirmation of the largest dips in observatory archival media. As such, we have identified the following historical dates in which one of \textit{Kepler}'s deepest dip (D1519 approx. 20\% and D1568 approx. 8\%) might be observed on such plates (see Tables 2 and 3).

The \textit{Kepler} mission made observations of KIC 8462852 every 29.4 minutes from May 1, 2009 up until May 11, 2013 when it experienced a fatal mechanical failure involving a second reaction wheel. Therefore, \textit{Kepler} observations covered a total of 1,471 days, a period that is approximately 102 days less than the hypothesized 1574-day periodicity. Consequently, there is a 102-day period in which no \textit{Kepler}-based predictions can be made. 
 
It should be noted that at the time of this paper, no space-borne missions are collecting daily observations of KIC 8462852. Future observations should strive to obtain measurements with night-to-night differential photometry better than 1\%. Consequently, upcoming predictions will need to be monitored closely since they are very small (0.2-0.5\%) and might be challenging to detect. On the other hand, a major dip or an attenuated/altered variant might be expected during a hypothesized return of D792 (a 16\% dip) on October 17, 2019 assuming this transit is on the same 1574-day orbit. 

\begin{table}
\begin{center}
\caption{Testable Predictions - A summary of the predicted recurrence of future dips based on the approximate 1574-day periodicity. It is important to note that only dips associated with the \textit{Kepler} light curve between 1500 and 1590 can be predicted with high confidence. The remaining dip predictions assumes all objects are on the same 1574-day orbit.\label{tbl2}}
\begin{tabular}{crrc}
\tableline
\tableline
Dip & Name & Depth & Date of next occult  \\
\tableline
1 & D140 & 0.5\% & 3-Jan-2018 \\
2 & D260 & 0.5\% & 3-May-2018 \\
3 & D359 & 0.2\% & 9-Aug-2018 \\
4 & D425 & 0.2\% & 14-Oct-2018 \\
5 & D792 & 16.0\% & 17-Oct-2019 \\
6 & D1205 & 0.4\% & 3-Dec-2020 \\
7 & D1487 & 2.0\% & 10-Sep-2021 \\
8 & D1519 & 21.0\% & 13-Oct-2021 \\
9 & D1540 & 3.0\% & 3-Nov-2021 \\
10 & 1568 & 8.0\% & 1-Dec-2021 \\
11 & Angkor & 3.2\% & 22-Jan-2022 \\
\tableline
\end{tabular}


\end{center}
\end{table}

\section{Discussion}

We have proposed a 1574-day periodicity that explains the striking similarity between the complex sequence of dimming events of KIC8462852 observed during the last two quarters of the \textit{Kepler} mission and dips observed more recently from the ground. This result is certainly subject to the poor sampling due to limited number of orbital observations. However, if this hypothesized 1574-day periodicity is confirmed by further observations, we can calculate the transiting bodies’ orbit radius assuming it is circular. Such a calculation reveals an interesting implication in that orbiting material, causing the complex light curve, would be located at approximately 2.983 AU. This distance is within the habitable zone confined to 2.174 and 3.132 AU, based on an absolute magnitude of 3.08 and a bolometric correction of -0.15 required for an F-star such as KIC 8462852. If this 1574-day periodicity is verified, one major challenge will be to understand how circumstellar material located at 3 AU from the star can result in such a complex sequence of dimming events. It is however worth pointing out that astronomer Bruce Gary mentioned on his webpage (http://www.brucegary.net/ts6/) that he may have detected a small (~1\%) dimming event on May 3, 2018. This date coincides exactly with the expected return of Kepler D260 x2 (see Table 2).

While we eagerly await what future observations will bring us, we can already look back at historical results using archived observatory plates.  To that end, Castelaz et al., examination of KIC 8462852 historical photographic plates archived at the Maria Mitchell Observatory provides evidence in support of a 1574.4-day periodicity.  In their paper, \citet{castelaz2018} identified five (5) possible short term dimming events / dips. As in all observatory archives, there are sporadic historical observations of KIC 8462852 (some not occurring for weeks, months, etc., between observations).  However, the identification of five dips presents an excellent opportunity to compare against the Kepler and 2017 LCO observations using a 1574.4 day periodicity.  Out of the five historical dips identified, only two of them would have fallen within the same window of time using Kepler D1487 – D1568 and May 2017 – September 2017. For these two dips that did fit our window, our question was, do they align precisely to any of the Kepler and LCO dips using a 1574.4-periodicity? We found that both of the two Castelaz et al. dips precisely match to the day:
  
\begin{itemize}
    \item Skara Brae minus 1574 = Kepler D1568
    \item Kepler D1568 minus (1574.4 X 9) = Match 1: October 22, 1978
    \item July 2017 dip minus 1574 = Kepler D1542
    \item Kepler D1542 minus (1574.4 X 18) = Match 2: August 21, 1935
\end{itemize}

Castelaz's et al., other three dips (using 1574.4) would have fallen outside of the 2017 events historically. This may lend support that other Kepler dips (beyond D1487 - D1568) are on a different orbit, although this point is completely unclear at this time. That being said, the July 16, 1966 dip is 30 days of D260 and the October 1980 dip is 80 days of D792. It is worth noting that Castelaz et al. used eight comparison stars and had a mean uncertainty 0.07 magnitude and the 1978 dip dimmed by at least 10\% increasing this sigma result. Furthermore, first identified by Hippke et al., there is a second observation of the October 22, 1978 dip by another observatory (Sonneberg, see Figure 7). Hippke et al., examined historical plate data from the DASCH digital archive (http://dasch.rc.fas.harvard.edu/project.php,  managed by the Harvard College Observatory, HCO) and from Sonneberg, and Sternberg observatories. Specifically, we reviewed the brightness magnitude (as found by Hippke et al.) of KIC 8462852 on the dates as found within Tables 3 and 4, which are the dates that we would expect to find a dip using a 1574-day and 1574.4-day periodicity and subtracting from both D1519 and D1568. Using these three observatories, there were two observations made of this star during these calculated dates:  October 24, 1978 at 8\% (Sonneberg) and April 30, 1944 at 6\% (DASCH).

The Sonneberg finding is an intriguing observation because it fits the same data found by Castelaz et al., using completely different plates (Maria Mitchell Observatory). The Sonneberg finding was first identified by \citet{hippke2017}, and is a potential 8\% dip occurring around October 24, 1978 (see Table 3, epoch 8 at 1574.4-day periodicity). Given the two separate observatories, and quality of plates, we believe the 1978 dip to be a multi-sigma detection.

\begin{figure}[h]
  \begin{center}
          \includegraphics[width=80mm] 
          {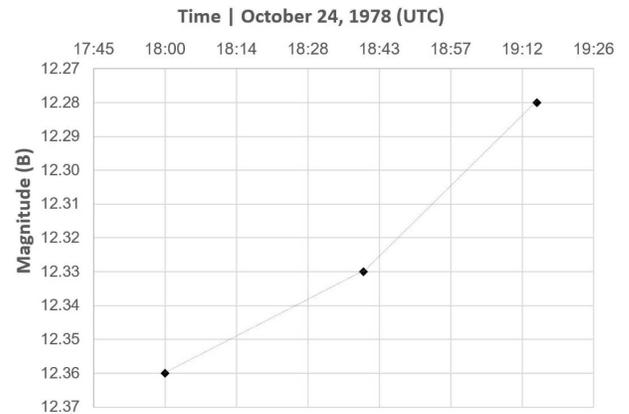}
    \caption{Graph representing depression of light for KIC 8462852 on three separate plates from October 24, 1978. This finding of a potential dip was made by Hippke et al. and documented within the paper, ``Sonneberg plate photometry for Boyajian's star in two passbands,'' The Astrophysical Journal, 2017.}\label{dip_hippke}
  \end{center}
\end{figure}

But what about the April 30, 1944 plate within the DASCH archive?  Once again using a 1574.4-day periodicity, we find that D1568 and Skara Brae should have been observable during this exact date in 1944.  DASCH records show that indeed this star did dim ~0.07 magnitude (approximately 6\%).  However, while interesting in itself as yet another positive result, there was only one plate and the plate quality is poor. Nonetheless, we are adding this finding within this discussion as it demonstrates our exhaustive effort to determine historical results.     

In the original ``Where's the Flux'' paper \citet{boyajian2016}, noted the apparent occurrence of dip separations that are multiples of 48.4, in some cases at half phase, or 24.2 days. For example, the separation between dips D792 and D1519 is approximately 15 periods of 48.6 days. We note that the proposed period of 1574 days is equivalent to 65 even periods of 24.2 days, even if the relevance, if any, of this ratio value remains to be determined.  Still, this is yet again another apparent result involving 24.2 (1574.0 / 24.2 = 65.0).  
\\

Based on this hypothesized periodicity, we provide testable prediction (historical and future) dates of possible dimming events. Should these predictions be verified, this would not only validate this periodicity hypothesis, but it would constitute a new set of extraordinary observations relating to this peculiar star. It would be a significant step forward in understanding the underlying mechanisms behind these dimming events. 

\begin{table}[h]
\begin{center}
\caption{Historical Occurrences of D1568 - The calculated timing of potential past occurrences of D1568 based on an approximate 1574-day periodicity.
\label{tbl3}}
\begin{tabular}{ccc}
\tableline
\tableline
Epoch (-D1568) & 1574-day periodicity & 1574.4-day periodicity  \\
\tableline
1 & 9-Aug-2017 & 9-Aug-2017 \\
0 & 18-Apr-2013 & 18-Apr-2013 \\
-1 & 26-Dec-2008 & 26-Dec-2008 \\
-2 & 4-Sep-2004 & 3-Sep-2004 \\
-3 & 14-May-2000 & 13-May-2000 \\
-4 & 22-Jan-1996 & 20-Jan-1996 \\
-5 & 1-Oct-1991 & 29-Sep-1991 \\
-6 & 10-Jun-1987 & 8-Jun-1987 \\
-7 & 17-Feb-1983 & 14-Feb-1983 \\
-8 & 27-Oct-1978 & 24-Oct-1978 \\
-9 & 6-Jul-1974 & 2-Jul-1974 \\
-10 & 15-Mar-1970 & 11-Mar-1970 \\
-11 & 22-Nov-1965 & 18-Nov-1965 \\
-12 & 1-Aug-1961 & 27-Jul-1961 \\
-13 & 10-Apr-1957 & 5-Apr-1957 \\
-14 & 18-Dec-1952 & 12-Dec-1952 \\
-15 & 27-Aug-1948 & 21-Aug-1948 \\
-16 & 6-May-1944 & 30-Apr-1944 \\
-17 & 14-Jan-1940 & 7-Jan-1940 \\
-18 & 23-Sep-1935 & 16-Sep-1935 \\
-19 & 2-Jun-1931 & 25-May-1931 \\
-20 & 9-Feb-1927 & 1-Feb-1927 \\
\tableline
\end{tabular}
\end{center}
\end{table}

\begin{table}[h]
\begin{center}
\caption{Historical Occurrences of D1519 - The calculated timing of potential past occurrences of D1519 based on an approximate 1574-day periodicity.\label{tbl4}}
\begin{tabular}{ccc}
\tableline
\tableline
Epoch (-D1519) & 1574-day periodicity & 1574.4-day periodicity  \\
\tableline
1 & 21-Jun-2017 & 21-Jun-2017 \\
0 & 28-Feb-2013 & 28-Feb-2013 \\
-1 & 7-Nov-2008 & 7-Nov-2008 \\
-2 & 17-Jul-2004 & 16-Jul-2004 \\
-3 & 26-Mar-2000 & 25-Mar-2000 \\
-4 & 4-Dec-1995 & 2-Dec-1995 \\
-5 & 13-Aug-1991 & 11-Aug-1991 \\
-6 & 22-Apr-1987 & 20-Apr-1987 \\
-7 & 30-Dec-1982 & 27-Dec-1982 \\
-8 & 8-Sep-1978 & 5-Sep-1978 \\
-9 & 18-May-1974 & 14-May-1974 \\
-10 & 25-Jan-1970 & 21-Jan-1970 \\
-11 & 4-Oct-1965 & 30-Sep-1965 \\
-12 & 13-Jun-1961 & 8-Jun-1961 \\
-13 & 20-Feb-1957 & 15-Feb-1957 \\
-14 & 30-Oct-1952 & 24-Oct-1952 \\
-15 & 9-Jul-1948 & 3-Jul-1948 \\
-16 & 18-Mar-1944 & 12-Mar-1944 \\
-17 & 26-Nov-1939 & 19-Nov-1939 \\
-18 & 5-Aug-1935 & 29-Jul-1935 \\
-19 & 14-Apr-1931 & 6-Apr-1931 \\
-20 & 22-Dec-1926 & 14-Dec-1926 \\
\tableline
\end{tabular}
\end{center}
\end{table}

\section{Conclusion}

On the basis of several sources of photometric data for KIC8462852 covering the longest epoch possible, we have provided support for a 1574-day periodicity of the complex dimming events that have been observed in the light curve by the \textit{Kepler} mission and ground-based telescopes. Based on this periodicity, we formulated testable predictions regarding the exact timing of future events. If confirmed, this periodicity would constrain further the mechanisms at play in this unique and fascinating solar system, notably involving circumstellar material orbiting the star in its habitable zone at approximately 3 AU.

\section{Acknowledgments}

The authors thank Dr. Tabetha Boyajian for her assistance in both providing access to LCO network data regarding KIC8462852, and invaluable commentary on this manuscript. We'd also like to point out that this paper would not have been possible without her Kick-Starter initiative which ultimately resulted in LCO observations. 

We also thank Jose Solorzano for his useful cross correlation suggestions and manuscript comments.  Finally, we thank Michael Hippke for his comments on the paper and for providing archival plate data for Sonneberg, DASCH, and Sternberg.  

Finally, the authors acknowledge the support of the DASCH project at Harvard by partial support from NSF grants AST-0407380, AST-0909073, and AST-1313370.




\clearpage

\end{document}